\documentclass[aps,pre,reprint,superscriptaddress]{revtex4-1}
\usepackage{graphicx}
\usepackage{amsmath}
\usepackage{amssymb}
\usepackage{bm}
\usepackage{physics}
\usepackage{subfigure}
\usepackage[colorlinks=true,linkcolor=blue,urlcolor=blue,citecolor=blue]{hyperref}
\usepackage{times}
\begin{document}

\title{Non-Markovian effects in stochastic resonance in a two level system}
\author{Ruofan Chen}
\affiliation{College of Physics and Electronic Engineering, and Center for Computational Sciences, Sichuan Normal University, Chengdu 610068, China}
\author{Xiansong Xu}
\affiliation{Science and Math Cluster, Singapore University of Technology and Design, 8 Somapah Road, Singapore 487372}
\date{\today}

\begin{abstract}
  Stochastic resonance is a phenomenon where the response signal to
  external driving is enhanced by environment noise. In quantum
  regime, the effect of environment is often intrinsically
  non-Markovian. Due to the combination of such non-Markovian quantum
  noise and external driving force, it is difficult to evaluate the
  correlation function and hence the power spectrum. Nevertheless, a
  recently developed algorithm, which is called time-evolving matrix
  product operators (TEMPO), and its extensions provide an efficient
  and numerically exact approach for this task. Using TEMPO we
  investigate non-Markovian effects in quantum stochastic resonance in
  a two level system. The periodic signal and the time-averaged
  asymptotic correlation function, along with the power spectrum, are
  calculated. From the power spectrum the signal-to-noise ratio is
  evaluated. It is shown that both signal strength and signal-to-noise
  ratio are enhanced by non-Markovian effects, which indicates the
  importance of non-Markovian effects in quantum stochastic
  resonance. In addition, we show that the non-Markovian effects can
  shift the peak position of the background noise power spectrum.
\end{abstract}
\maketitle

\section{Introduction}
Stochastic resonance (SR) is phenomenon for which the response of the
system to external driving is enhanced by noise, which was first
proposed by Benzi \textit{et al.} \cite{benzi1981-the}.  Since then SR
has continuously attracted considerable attention over several
decades. The first experimental verification of such phenomenon was
obtained by Fauve and Heslot \cite{fauve1983-stochastic}, who studied
the noise-induced transition process in a bistable system. Another key
experiment in this field is the observation of SR in an optical
device, the bidirectional ring laser, by McNamara \textit{et al.}
\cite{mcnamara1988-observation}.  The concept of SR was extended to
the quantum regime, which is referred as quantum stochastic resonance
(QSR), by L\"{o}fstedt and Coppersmith \cite{loefstedt1994-quantum} by
measuring conductance fluctuations in mesoscopic metals. A control
scheme for SR using a Schmitt trigger is demonstrated by Gammaitoni
\textit{et al.} \cite{gammaitoni1999-controlling}. Recently, QSR is
demonstrated in the a.c.-driven charging and discharging of single
electron on a quantum dot
\cite{wagner2019-quantum,hussein2020-spectral} and in individual Fe
atoms \cite{haenze2021-quantum}.

The theoretical investigation for SR in terms of periodically driven
dissipative system starts several decades ago, and comprehensive
reviews can be found in
Refs. \cite{mcnamara1989-theory,jung1993-periodically,grifoni1998-driven,gammaitoni1998-stochasticresonance}.
When the quantum coherence is suppressed, both quantum and classical
SR can be well described by the classical rate equation approach
\cite{mcnamara1989-theory}. In deep quantum regime, qualitative new
features arise \cite{grifoni1996-quantum,grifoni1996-coherent}. This
regime has been investigated semiclassically by Grifoni \textit{et
  al.}  \cite{grifoni1996-quantum} and numerically by Makarov and
Makri \cite{makarov1995-stochastic,makarov1995-control}. The QSR in a system driven by weak
signal and white noise was studied by Joshi
\cite{joshi2008-stochastic}.

The non-Markovian transient property of QSR is difficult to evaluate
due to the interplay of dissipative and external driving.  A
numerically exact method known as quasi-adiabatic propagator path
integral (QUAPI) method
\cite{makarov1995-stochastic,makarov1995-control}, which fully takes
the non-Markovian effects into consideration, is meant to be suitable
for such task. However, the computational cost of QUAPI grows
exponentially with size of the system Hilbert space and memory length,
therefore although numerically exact, QUAPI can become highly
inefficient or even infeasible under certain circumstances. Due to
this limitation, the correlation function and the corresponding power
spectrum, which is the quantity of fundamental interest in QSR study,
are not suitable to be evaluated by bare QUAPI algorithm.

Recently, Strathearn \textit{et al.} \cite{strathearn2018-efficient}
show that QUAPI method can be represented in the framework of matrix
product states (MPS)
\cite{schollwoeck2011-density,orus2014-practical}. The standard MPS
compression algorithm is applicable in this framework, and thus they
obtain an efficient and numerically exact method which is called
time-evolving matrix product operators (TEMPO). Later J{\o}rgensen and
Pollock \cite{joergensen2019-exploiting} relate TEMPO to process
tensor to motivate an efficient and numerically exact algorithm for
simulation of correlation functions in undriven open systems. The
TEMPO method is also modified for repeated computation of various sets
of parameters by Fux \textit{et al.} \cite{fux2021-efficient}.

In this article, we employ TEMPO to study the non-Markovian effects in
QSR in an open two level system which is periodically driven. The
periodic signal is evaluated for which the signal strength shows a
maximum as noise level increases, which is a sign of QSR. In the
Markovian limit, the signal almost vanishes, while the signal
reappears when non-Markovian effects are included. The signal strength
increases with increasing non-Markovianity, which shows that the
non-Markovian effects plays an essential role in QSR.

The asymptotic correlation function are evaluated with different
initial time. Unlike the periodic signal, whose strength increases
with increasing non-Markovianity, the amplitude of asymptotic
correlation function can remain very small with certain initial
time. The Fourier transform of the time-averaged asymptotic
correlation gives the power spectrum, from which the background noise
power and signal-to-noise ratio (SNR) are obtained. The SNR is also
enhanced by non-Markovian effects. The peak position of background
noise power can be shifted by the non-Markovian effects, which makes
that the maximum of signal strength and SNR appear at different noise
levels.

This article is organized as follows. The introduction of the model
and method are given in Sec. \ref{sec:model} and \ref{sec:method},
respectively. The non-Markovian effects on observable and correlation
function are discussed in Sec. \ref{sec:observable} and
\ref{sec:correlation-function}, respectively. Section
\ref{sec:signal-to-noise-ratio} gives the result of signal-to-noise
ratio. Finally a conclusion is given in Sec. \ref{sec:conclusions}.

\section{Model}
\label{sec:model}
Bistable system is the simplest and also the most widely used model to
study stochastic resonance. Such a bistable system can be effectively
described by a two level system, when external driving is present the
Hamiltonian can be written as
\begin{equation}
  H_{\mathrm{S}}(t)=\frac{\Delta}{2}\sigma_x+\frac{E}{2}\sigma_z\cos\Omega t,
\end{equation}
where $\sigma_z$ and $\sigma_x$ are Pauli matrices, and the
eigenstates of $\sigma_z$ are the basis states in a localized
representation. Here $\Delta$ gives the tunneling amplitude between
two levels, $E$ is the strength of external driving and $\Omega$ is
driving frequency.

Here we consider a Caldeira-Leggett type model
\cite{caldeira1983-path,caldeira1983-quantum} for which the bath
Hamiltonian and the system-bath coupling are
\begin{equation}
  H_{\mathrm{B}}=\sum_k\omega_kb_k^{\dag}b_k,\quad
  H_{\mathrm{SB}}=\sigma_z\sum_kV_k(b_k^{\dag}+b),
\end{equation}
where the operator $b_k^{\dag}$ ($b_k$) creates (annihilates) a boson
in state $k$ with frequency $\omega_k$. The total Hamiltonian is
\begin{equation}
  H=H_{\mathrm{S}}(t)+H_{\mathrm{B}}+H_{\mathrm{SB}}.
\end{equation}
This model is usually referred as spin-boson model
\cite{leggett1987-dynamics,weiss1993-quantum}. The bath is
characterized by a spectral function $J(\omega)$ and in this article
we choose Ohmic dissipation for which
\begin{equation}
  J(\omega)=\lambda\omega e^{-\omega/\omega_c},
\end{equation}
where $\lambda$ is the coupling strength parameter and $\omega_c$ is
the cutoff frequency. If the bath is in thermal equilibrium state then
the bath autocorrelation function is written in terms of $J(\omega)$
as
\begin{equation}
  \label{eq:autocorrelation}
  \alpha(t)=\int_0^{\infty} J(\omega)\qty[\coth(\frac{\omega}{2T})\cos\omega t-i\sin\omega t]\dd{\omega},
\end{equation}
where $T$ is the temperature. The noise level is determined by the
coupling strength and the temperature. Throughout this article we set
$\hbar=k_B=1$ and use dimensionless quantities.

The relevant theoretical quantity describing the dissipative dynamics
is the expectation value of observable $\expval{\sigma_z(t)}$. Under
periodic driving, this quantity shows coherent oscillations in steady
state whose amplitude gives the periodic signal strength. The signal
strength shows a maximum with respect to noise level
\cite{makarov1995-stochastic,makarov1995-control}, which is a sign of
QSR.

Sometimes the correlation function is of more concern. The symmetrized
correlation function is defined
\begin{equation}
  C(t_1,t_2)=\frac{1}{2}\expval{\sigma_z(t_1)\sigma_z(t_2)+\sigma_z(t_2)\sigma_z(t_1)}.
\end{equation}
Under this definition, the correlation function is symmetrized in the
sense that $C(t_1,t_2)=C(t_2,t_1)$, and we can write
\begin{equation}
  C(t_1,t_2)=\begin{cases}
    \Re\expval{\sigma_z(t_1)\sigma_z(t_2)},& t_2\ge t_1;\\
    \Re\expval{\sigma_z(t_2)\sigma_z(t_1)},& t_1\ge t_2.\\
    \end{cases}
\end{equation}
It means the symmetrized correlation function $C(t_1,t_2)$ is a real
quantity and the knowledge of the correlation function
$\expval{\sigma_z(t_1)\sigma_z(t_2)}$ for $t_2\ge t_1$ is enough to
obtain whole $C(t_1,t_2)$. Therefore we focus on $C(t_0,t_0+t)$ for
$t\ge0$ situation. The time-averaged, asymptotic (i.e.,
$t_0\to\infty$) symmetrized correlation function is defined as
\begin{equation}
  \label{eq:time-averaged-C}
  \bar{C}(t)=\lim_{t_0\to\infty}\frac{\omega}{2\pi}
  \int_0^{\frac{2\pi}{\omega}}C(t_0,t_0+t)\dd{t_0}.
\end{equation}
The quantity of experimental interest for QSR is the power spectrum
\cite{mcnamara1989-theory,grifoni1998-driven}, which is defined
one-sided, i.e., for positive $\omega$ only, as
\begin{equation}
  \begin{split}
    \bar{S}(\omega)&=\frac{1}{2}\int_{-\infty}^{\infty}\bar{C}(t)(e^{i\omega t}+e^{-i\omega t})\dd{t}\\
    &=\int_{-\infty}^{\infty}\bar{C}(t)\cos\omega t\dd{t}.
  \end{split}
\end{equation}

\section{Method}
\label{sec:method}
\subsection{Quasi-Adiabatic Propagator Path Integral}
Here we give a brief review of the QUAPI and TEMPO method. Let
$\rho(t)$ denote the total density matrix, then the time evolution of
$\rho$ is given by
\begin{equation}
  \rho(t)=U(t,0)\rho(0)U^{\dag}(t,0),
\end{equation}
where
\begin{equation}
  U(t,t_0)=\mathrm{T}\exp[-i\int_{t_0}^tH(\tau)\dd{\tau}].
\end{equation}
Here $\mathrm{T}$ denotes the chronological ordering operator.  The
reduced density matrix is defined as
$\rho_\mathrm{S}(t)=\Tr_{\mathrm{B}}[\rho(t)]$, where
$\Tr_{\mathrm{B}}$ denote the trace over the bath degrees of
freedom. Let $s$ denote the eigenvalue of $\sigma_z$, then the element
of the reduced density matrix can be written as
\begin{equation}
\rho_{\mathrm{S}}(s'',s';t)=\Tr_{\mathrm{B}}\mel{s''}{U(t,0)\rho(0)U^{\dag}(t,0)}{s'}.
\end{equation}
If at initial time $t=0$ the total density
matrix is in product state for which
\begin{equation}
  \rho(0)=\rho_{\mathrm{S}}(0)\otimes\rho_{\mathrm{B}},
\end{equation}
then the above expression can be written in path integral form by
splitting the evolution into $N$ pieces for which $\delta t=t/N$
with $N\to\infty$. Relabeling $s'',s'$ as $s_N^+,s_N^-$ yields
\begin{equation}
  \label{eq:rho-calculation}
  \rho_{\mathrm{S}}(s_N^{\pm})=
  \sum_{s_0^{\pm},\ldots,s_{N-1}^{\pm}}F(s_0^{\pm},\ldots,s_N^{\pm}),
\end{equation}
where $\rho_{\mathrm{S}}(s_N^{\pm})$ denotes
$\mel{s_N^+}{\rho_{\mathrm{S}}(t)}{s_N^-}$ and
\begin{equation}
  \label{eq:tilde-rho}
  F(s_0^{\pm},\ldots,s_N^\pm)=
  K(s_0^{\pm},\ldots,s_N^{\pm})I(s_0^{\pm},\ldots,s_N^{\pm}).
\end{equation}
Here $K(s_0^{\pm},\ldots,s_N^{\pm})$ is the bare system propagating
tensor for which
\begin{equation}
  K(s_0^{\pm},\ldots,s_N^{\pm})=\rho_{\mathrm{S}}(s_0^\pm)\tilde{K}(s_0^{\pm},s_1^{\pm})
  \cdots \tilde{K}(s_{N-1}^{\pm},s_N^{\pm}),
\end{equation}
where ($t_k=k\delta t$) 
\begin{equation}
  \label{eq:K}
  \begin{split}
    \tilde{K}(s_{k-1}^{\pm},s_k^{\pm})=&\mel{s_k^+}{U(t_k,t_{k-1})}{s_{k-1}^+}\\
    &\times\mel{s_{k-1}^-}{U^{\dag}(t_k,t_{k-1})}{s_k^-}.
  \end{split}
\end{equation}

In the continuum limit $\delta t\to0$, the collections of
$(s_0^+,\ldots,s_N^+)$ and $(s_0^-,\ldots,s_N^-)$ can be regarded as a
forward path $s^+(t')$ and a backward path $s^-(t')$ from 0 to $t$. If
at initial time the bath is in thermal equilibrium state that
$\rho_{\mathrm{B}}=e^{-H_{\mathrm{B}}/T}$, then the influence
functional $I[s^{\pm}(t)]$ can be written as
\cite{feynman1963-the,feynman1965-quantum,caldeira1983-path}
\begin{equation}
  e^{-\int_0^t\dd{t'}\int_0^{t'}\dd{t''}
    [s^+(t')-s^-(t')][\alpha(t'-t'')s^+(t'')-\alpha^{*}(t'-t'')s^-(t'')]},
\end{equation}
where $\alpha(t)$ is the autocorrelation function given by
\eqref{eq:autocorrelation}. When employing finite $\delta t$
approximation, this influence functional can be discretized as
\begin{equation}
  \label{eq:influence-functional}
  I(s_0^{\pm},\ldots,s_N^{\pm})=\exp[-\sum_{j=0}^N\sum_{k=0}^j\phi_{jk}],
\end{equation}
where
\begin{equation}
  \phi_{jk}=(s_j^+-s_j^-)(\eta_{jk}s_k^+-\eta_{jk}^{*}s_k^-).
\end{equation}
Here the form of $\eta_{jk}$ depends on choice of discretization
scheme \cite{dattani2012-analytic}.

The discretized influence functional \eqref{eq:influence-functional}
is a tensor which can be decomposed as
\begin{equation}
  \label{eq:decomposition}
  \begin{split}
    I(s_0^{\pm},\ldots,s_N^{\pm})=&\prod_{k=0}^NI_0(s_k^{\pm},s_k^{\pm})\prod_{k=0}^{N-1}I_1(s_k^{\pm},s_{k+1}^{\pm})\cdots\\
    &\times\prod_{k=0}^{N-\Delta k}I_{\Delta k}(s_k^{\pm},s_{k+\Delta k}^{\pm})\cdots\\
    &\times I_N(s_0^{\pm},s_N^{\pm}),
  \end{split}
\end{equation}
where
\begin{equation}
  I_{\Delta k}(s_k^{\pm},s_{k+\Delta k}^{\pm})=e^{-\phi_{k+\Delta k,k}}.
\end{equation}
The key idea of QUAPI method is that non-locality of $\eta_{jk}$ drops
off as $\Delta k$ increases, then $\eta_{k+\Delta k,k}$ can be
neglected when $\Delta k$ is greater than a certain positive integer
$N_s$. Therefore the influence functional \eqref{eq:decomposition}
can be truncated as
\begin{equation}
  \label{eq:decomposition-truncation}
  \begin{split}
    I(s_0^{\pm},\ldots,s_N^{\pm})=&\prod_{k=0}^NI_0(s_k^{\pm},s_k^{\pm})\prod_{k=0}^{N-1}I_1(s_k^{\pm},s_{k+1}^{\pm})\cdots\\
    &\times\prod_{k=0}^{N-\Delta k}I_{N-\Delta k}(s_k^{\pm},s_{k+\Delta k}^{\pm})\cdots\\
    &\times\prod_{k=0}^{N-N_s}I_{N_s}(s_k^{\pm},s_{k+N_s}^{\pm}).
\end{split}
\end{equation}
Now define a tensor $A(s_0^{\pm},\ldots,s_{k+1}^{\pm})$ as
\begin{equation}
  \label{eq:A-tensor}
  \tilde{K}(s_k^{\pm},s_{k+1}^{\pm})
  \prod_{\Delta k=0}^{\min(k+1,N_s)}I_{\Delta k}(s_{k+1-\Delta k}^{\pm},s_{k+1}^{\pm}),
\end{equation}
then there is a recursive relation for which
\begin{equation}
  \label{eq:recursive-relation}
  F(s_0^{\pm},\ldots,s_{k+1}^{\pm})=F(s_0^{\pm},\ldots,s_k^{\pm})A(s_0^{\pm},\ldots,s_{k+1}^{\pm}).
\end{equation}
Employing the above recursive relation iteratively we can get the
final $F(s_0^{\pm},\ldots,s_N^{\pm})$ from initial condition
\begin{equation}
  F(s_0^{\pm})=\rho_{\mathrm{S}}(s_0^{\pm})I_0(s_0^{\pm}).
\end{equation}

\subsection{Correlation Functions}
The formalism described above only deals with the dynamics of reduced
density matrix. It is, however, easy to be generalized for the
correlation function calculation.

The expectation value $\expval{\sigma_z(t')}$ can be obtained via
\begin{equation}
  \expval{\sigma_z(t')}=\Tr[U(t,0)\rho(0)U^{\dag}(t',0)\sigma_zU^{\dag}(t,t')],
\end{equation}
where $\Tr=\Tr_{\mathrm{S}}\Tr_{\mathrm{B}}$ is the trace over all
degrees of freedom with $\Tr_{\mathrm{S}}$ the trace over degrees of
freedom of the system. Suppose $t'=k\delta t$, we can define a
one-time ``correlated'' reduced density matrix as
\begin{equation}
  \tilde{\rho}_{\mathrm{S}}(s_N^{\pm};s_k^-)=\sum_{s_0^{\pm},\ldots,s_{N-1}^{\pm}}F(s_0^{\pm},\ldots,s_N^{\pm})s_k^-,
\end{equation}
then the expectation value can be written as
\begin{equation}
  \expval{\sigma_z(t')}=\Tr_{\mathrm{S}}[\tilde{\rho}_{\mathrm{S}}(s_N^{\pm},s_k^-)].
\end{equation}

As mentioned in Sec. \ref{sec:model}, we need to only calculate the
correlation function $\expval{\sigma_z(t_1)\sigma_z(t_2)}$ for
$t_2\ge t_1$.  In this case, $\expval{\sigma_z(t_1)\sigma_z(t_2)}$ can
be written as
\begin{equation}
  \begin{split}
    \Tr[U(t,0)\rho(0)U^{\dag}(t_1,0)\sigma_zU^{\dag}(t_2,t_1)\sigma_zU^{\dag}(t,t_2)].
  \end{split}
\end{equation}
Suppose $t_1=k_1\delta t$ and $t_2=k_2\delta t$, where $k_2\ge k_1$,
we can define a two-time ``correlated'' reduced density matrix as
\begin{equation}
  \tilde{\rho}_{\mathrm{S}}(s_N^{\pm};s_{k_1}^-,s_{k_2}^-)=\sum_{s_0^{\pm},\ldots,s_{N-1}^{\pm}}F(s_0^{\pm},\ldots,s_N^{\pm})s_{k_1}^-s_{k_2}^-,
\end{equation}
and the correlation function is obtained via
\begin{equation}
  \label{eq:correlation-calculation}
  \expval{\sigma_z(t_1)\sigma_z(t_2)}=\Tr_{\mathrm{S}}[\tilde{\rho}_{\mathrm{S}}(s_N^{\pm};s_{k_1}^-,s_{k_2}^-)].
\end{equation}

\subsection{Time-Evolving Matrix Product Operators}
So far we have discussed the basic framework of calculating the
correlation function in a driven spin-boson model. However,
$F(s_0^{\pm},\ldots,s_N^{\pm})$ is a tensor of rank $2(N+1)$ for which
a space with size proportional to $2^{2(N+1)}$ is needed to store it.
In practical calculation it is very difficult to handle such tensor
directly unless $N$ is fairly small.

In original QUAPI
\cite{makarov1993-tunneling,makarov1994-path,makri1995-numerical}, an
iterative tensor multiplication algorithm is employed and a tensor of
rank $2(N_s+1)$, rather than $2(N+1)$, is kept in track during the
time evolution process. This greatly reduce the space needed, but the
computational cost still scales exponentially with $N_s$.  The value
$N_s\delta t$ is supposed to cover the range of non-locality of
$\eta_{jk}$, then to ensure a small $N_s$ usually a relatively large
$\delta t$ is adopted, which may introduce relatively large Trotter
errors.

Recently, Strathearn \textit{et al.} \cite{strathearn2018-efficient}
showed that the tensor $F(s_0^{\pm},\ldots,s_N^{\pm})$ can be
naturally represented by matrix product states (MPS)
\cite{schollwoeck2011-density,orus2014-practical} and developed the
TEMPO algorithm. The main idea is that $F(s_0^{\pm},\ldots,s_N^{\pm})$
can be efficiently constructed via iterative application of matrix
product operator (MPO). Such iterative process is amenable to standard
MPS compression algorithm, and thus computational cost scales only
polynomially with $N_s$. This allows us to perform simulations to
large values of $N_s$, for instance, $N_s$ in
Ref. \cite{strathearn2018-efficient} is up to 200 which is impossible
to simulate without tensor compression algorithm. There is also
another approach for tensor network representation of discretized path
integral by Oshiyama \textit{et al.}
\cite{oshiyama2020-kibble,oshiyama2022-classical}.

Later J{\o}rgensen and Pollock \cite{joergensen2019-exploiting}
related the influence functional to process tensor via representing
the influence functional $I(s_0^{\pm},\ldots,s_N^{\pm})$ by MPS, they
use this connection to motivate a tensor network algorithm for
simulation of multiple time correlation functions. Fux \textit{et al.}
\cite{fux2021-efficient} modified TEMPO method for repeated
computation of various sets of parameters.

For clarity and simplicity, we abbreviate the index pair
$\{s_k^+,s_k^-\}$ as $s_k$. In this way the $2\times2$ reduced density
matrix is represented as a vector of $4$ elements. We also write $F$
as a superscripted tensor for which
\begin{equation}
  F^{s_0,\ldots,s_k}=F(s_0,\ldots,s_k).
\end{equation}

Define a $B$ tensor as
\begin{equation}
  B^{s_0,\ldots,s_{k+1}}_{r_0,\ldots,r_k}=\qty(\prod_{i=0}^k\delta_{s_kr_k})
  A(s_0,\ldots,s_{k+1}),
\end{equation}
where $A(s_0,\ldots,s_{k+1})$ is the tensor defined in
\eqref{eq:A-tensor}, then the recursive relation
\eqref{eq:recursive-relation} can be written in terms of Einstein
summation convention way as
\begin{equation}
  \label{eq:recursive-relation-tensor}
  F^{s_0,\ldots,s_{k+1}}=B^{s_0,\ldots,s_{k+1}}_{r_0,\ldots,r_k}F^{r_0,\ldots,r_k}.
\end{equation}

\begin{figure}[htbp]
  \centerline{\includegraphics[]{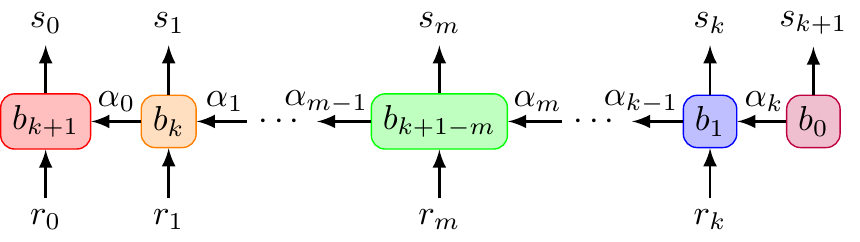}}
  \caption{Tensor network representation of the MPO
    $B^{s_0,\ldots,s_{k+1}}_{r_0,\ldots,r_k}$.}
  \label{fig:01}
\end{figure}

If $F^{s_0,\ldots,s_k}$ is represented as a MPS, then
$F^{s_0,\ldots,s_{k+1}}$ can keep the MPS structure if the $B$ tensor
is represented as a MPO. Then during the iterative process the
standard MPS compression algorithm can be applied such that the
required computational resource scales polynomially. The form meets the
requirement is (here Einstein summation convention still applies)
\begin{equation}
  \begin{split}
    B^{s_0,\ldots,s_{k+1}}_{r_0,\ldots,r_k}=&[b_{k+1}]^{s_0\alpha_0}_{r_0}
    [b_k]^{s_1\alpha_1}_{r_1\alpha_0}\cdots\\
    &[b_{k+1-m}]^{s_m\alpha_m}_{r_m\alpha_{m-1}}\cdots
    [b_1]^{s_k\alpha_k}_{r_k\alpha_{k-1}}[b_0]^{s_{k+1}}_{\alpha_k},
  \end{split}
\end{equation}
where the rank-$3$ tensor in the front is defined as
\begin{equation}
  [b_{k+1}]^{s_0\alpha_0}_{r_0}=I_{k+1}(s_0,\alpha_0)\delta^{s_0}_{r_0}.
\end{equation}
When $m<k$, the rank-$4$ tensors in the middle are defined as
\begin{equation}
  [b_{k+1-m}]^{s_m\alpha_m}_{r_m\alpha_{m-1}}=I_{k+1-m}(s_m,\alpha_m)
  \delta^{s_m}_{r_m}\delta^{\alpha_m}_{\alpha_{m-1}},
\end{equation}
and when $m=k$ we have
\begin{equation}
  [b_1]^{s_k\alpha_k}_{r_k\alpha_{k-1}}=\tilde{K}(s_k,\alpha_k)
  I_1(s_k,\alpha_k)\delta^{s_k}_{r_k}\delta^{\alpha_k}_{\alpha_{k-1}}.
\end{equation}
The last rank-$2$ tensor is defined as
\begin{equation}
  [b_0]^{s_{k+1}}_{\alpha_k}=I_0(s_{k+1},s_{k+1})\delta^{s_{k+1}}_{\alpha_k}.
\end{equation}
The tensor network representation of the MPO
$B^{s_0,\ldots,s_{k+1}}_{r_0,\ldots,r_k}$ is depicted in
Fig. \ref{fig:01}.

\begin{figure}[htbp]
  \centerline{\includegraphics[]{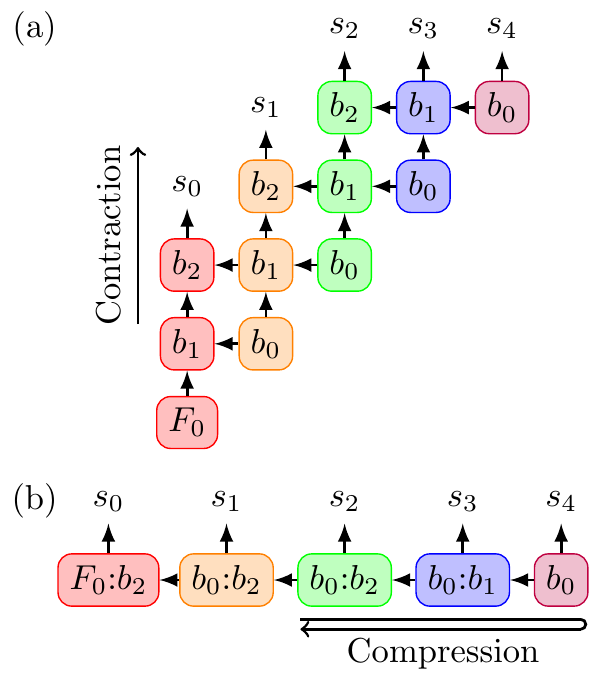}}
  \caption{(a) Tensor network representation of $F$ tensor of first
    five steps with truncation parameter $N_s=2$ before
    contraction. (b) After the contraction, the compression goes
    through $s_2$ to $s_4$ only.}
  \label{fig:02}
\end{figure}

The tensor network of $F$ tensor for first five steps is shown
graphically in Fig. \ref{fig:02}(a) with truncation parameter
$N_s=2$. Unlike the original QUAPI and TEMPO algorithm, here we do not
contract the indices beyond last $N_s+1$ steps for which $s_0$ and
$s_1$ are not summed out in the figure. That is, after the iterative
propagating process, we shall obtain $F$ as a tensor with indices from
$s_0$, rather than $s_{N-N_s+1}$, to $s_N$.

The way we arrange the tensor network shown in Fig. \ref{fig:02}(a) is
called nonlocal network boundary
\cite{joergensen2019-exploiting}. This nonlocal boundary choice is
supposed to be much more inefficient than the local boundary one if
there is no truncation $N_s$. The reason is that with the nonlocal
boundary choice, during each propagating process
\eqref{eq:recursive-relation-tensor} all time step indices are
affected and the MPS compression need to go through all the
indices. However, if $N$ is large the calculation would be
computationally costly even with the local boundary
condition. Therefore we still adopt the nonlocal boundary but with the
truncation $N_s$, and the compression only goes through the last
$N_s+1$ steps, as shown in Fig. \ref{fig:02}(b).

\section{Non-Markovian Effects on $\expval{\sigma_z(t)}$}
\label{sec:observable}
In this section we study the non-Markovian dynamics of the expectation
value $\sigma_z(t)$. Starting from an arbitrary initial state (here we
starts from $\expval{\sigma_z(0)}=1$), the reduced density matrix
would eventually reach a steady state where $\expval{\sigma_z(t)}$
oscillates with frequency $\Omega$. The coherently oscillating
$\expval{\sigma_z(t)}$ is just the periodic response signal.

In QUAPI and TEMPO algorithms, the non-Markovianity are controlled by
the truncation $N_s$.  If $N_s$ is large enough to cover the
non-locality of $\alpha(t)$ (or $\eta_{jk}$) then it is supposed to
capture all non-Markovian effects. On the other hand, if $N_s=1$ then
the dynamics is only relevant to one last time step and thus it gives
the Markovian result. In this article, we shall call the $N_s=1$ case
Markovian, as in Ref.  \cite{joergensen2019-exploiting}.

\begin{figure}[htbp]
  \centerline{\includegraphics[]{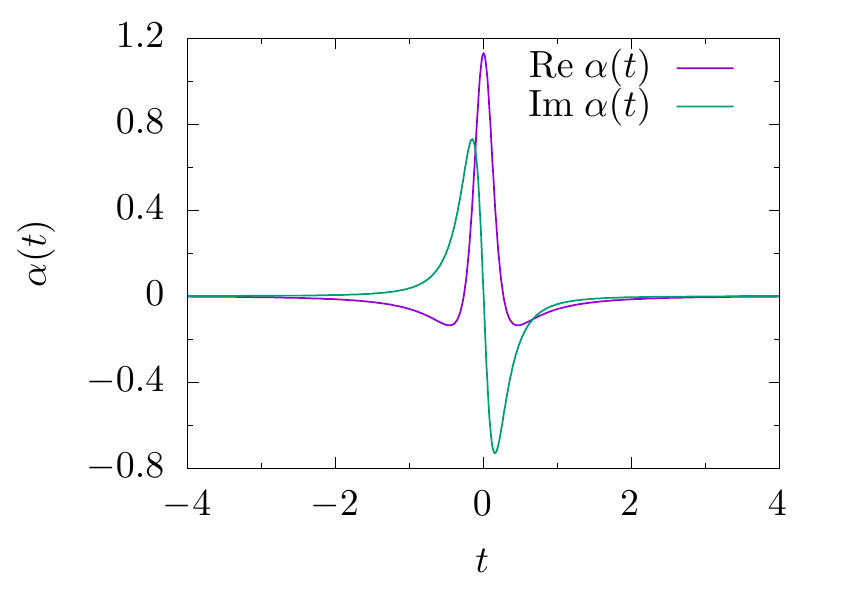}}
  \caption{The autocorrelation function $\alpha(t)$ with
    $\lambda=0.08$, $T=0.139$ and $\omega_c=3.75$.}
  \label{fig:03}
\end{figure}

A set of parameters which induces large amplitude oscillation can be
found in Refs. \cite{makarov1995-stochastic,makarov1995-control}. If
we set $\Delta=1$, then in our model they correspond to $\Omega=1$,
$E=\frac{1}{2}$, $\omega_c=3.75$, $T=0.139$ and $\lambda=0.08$. From
now on, we shall fix our parameters listed here except the coupling
strength $\lambda$. The autocorrelation function $\alpha(t)$ with
these parameters are shown in Fig. \ref{fig:03}. At $t=\pm4$, the
autocorrelation function already becomes very small.

\begin{figure}[htbp]
  \centerline{\includegraphics[]{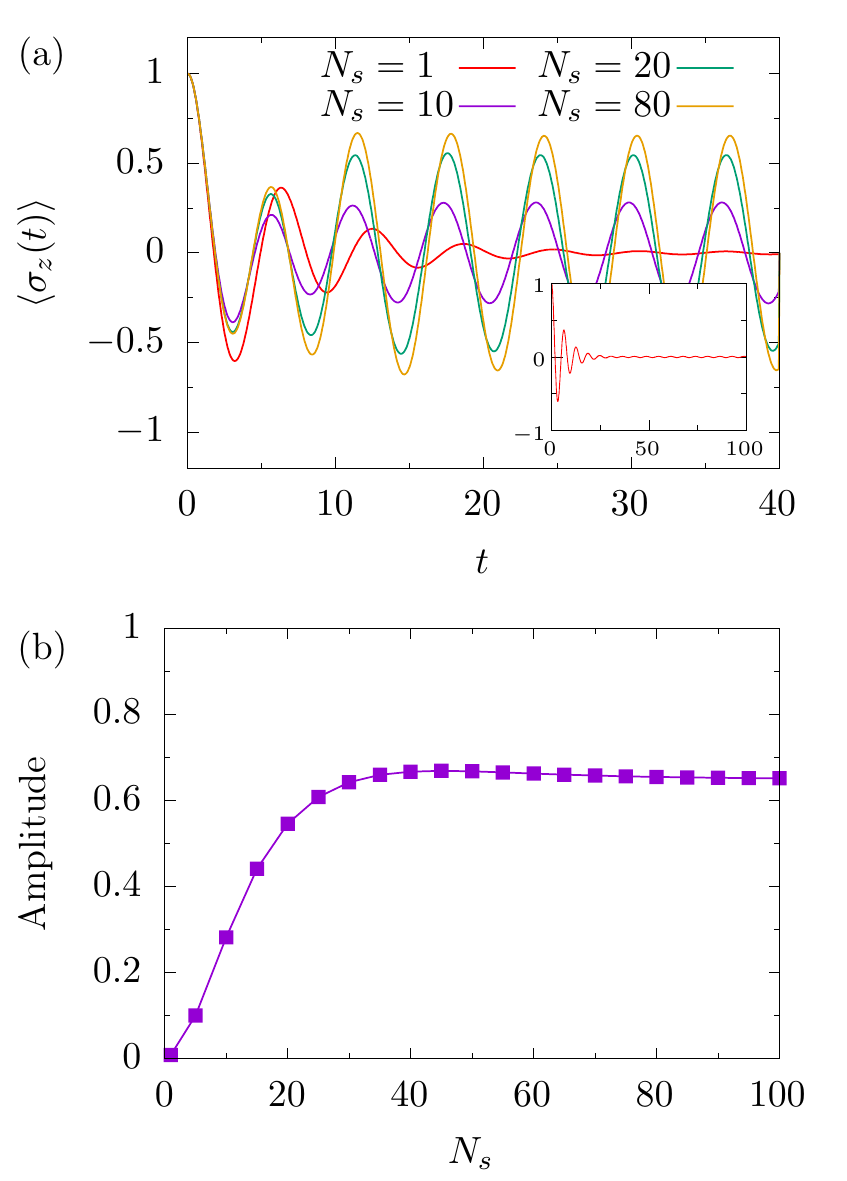}}
  \caption{(a) Some typical results of $\expval{\sigma_z(t)}$ with
    respect to different $N_s$. The inset shows a longer time scale
    for $N_s=1$, i.e., Markovian case. (b) The amplitudes of steady
    coherent oscillation with respect to different $N_s$. Here
    $\lambda=0.08$.}
  \label{fig:04}
\end{figure}

Now we want to cover the non-locality of $\alpha(t)$ shown in
Fig. \ref{fig:03}, i.e., $N_s\delta t\ge 4$. Typical simulations of
QUAPI are restricted to $N_s<20$
\cite{nalbach2011-iterative,thorwart2005-non}, and in fact when $N_s$
is greater than 10 it already become time consuming. Therefore the
time interval $\delta t$ is usually not less than $0.25$. By employing
TEMPO algorithm we go to $N_s=80$ in this article, and then $\delta t$
can reach a fairly small value $0.05$.

Some typical results of $\expval{\sigma_z(t)}$ with different $N_s$
are shown in Fig. \ref{fig:04}(a). It can be seen that the behavior of
Markovian $\expval{\sigma_z(t)}$ is qualitatively different from the
non-Markovian ones. When $N_s=1$ (the Markovian case),
$\expval{\sigma_z(t)}$ shows coherent decaying oscillation which
decays to a very small (almost zero) oscillation eventually (see the
inset of the figure). For non-Markovian cases, even with a not so
large $N_s=10$, $\expval{\sigma_z(t)}$ reaches steady states fast and
then oscillates coherently. When $N_s$ increases, the larger amplitude
coherent oscillation is induced as the more non-Markovian effects are
included. The amplitudes of coherent oscillation with respect to
different $N_s$ are shown in Fig. \ref{fig:04}(b). It can be seen that
the amplitude increases as $N_s$ increases and becomes stable when
$N_s\ge80$. This value corresponds $N_s\delta t\approx4$, at this
value most non-Markovian effects are captured.

\begin{figure}[htbp]
  \centerline{\includegraphics[]{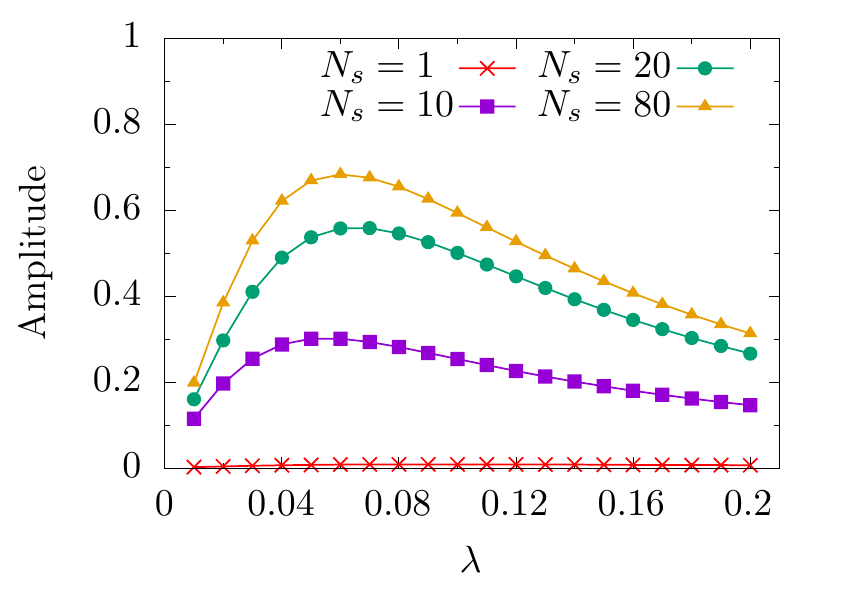}}
  \caption{Amplitudes of steady-state oscillation with respect to
    coupling strength $\lambda$ with different $N_s$.}
  \label{fig:05}
\end{figure}

Figure \ref{fig:05} shows the amplitudes of steady-state oscillation
with respect to coupling strength $\lambda$ with different $N_s$. For
non-Markovian case $N_s=80$, a pronounced maximum is
demonstrated. This is the sign of QSR phenomenon where the response of
a non-linear system to the external periodic driving is enhanced by
noise. Note that the maximum of amplitudes is at $\lambda\approx0.06$
rather than $\lambda=0.08$.

\section{Non-Markovian Effects on Correlation Function}
\label{sec:correlation-function}
In this section we study the non-Markovian effects on symmetrized
correlation function $C(t_0,t_0+t)$. As mentioned in
Sec. \ref{sec:model}, it is enough to calculate the $t\ge0$ case.

\begin{figure}[htbp]
  \centerline{\includegraphics[]{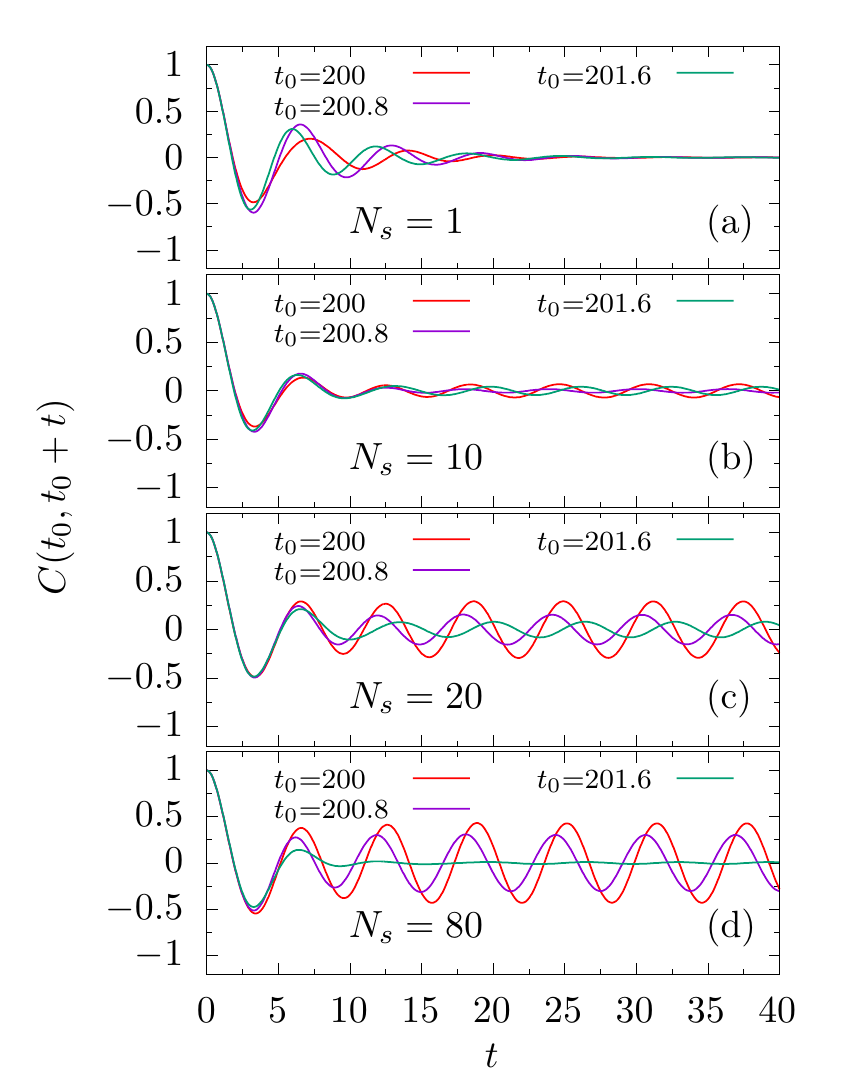}}
  \caption{Some typical symmetrized correlation function
    $C(t_0,t_0+t)$ with different $N_s$ and $t_0$. Here
    $\lambda=0.08$.}
\label{fig:06}
\end{figure}

The correlation function should be evaluated at steady state where
$\expval{\sigma_z(t)}$ is doing coherent oscillation, i.e., the time
$t_0$ should be large enough. It can be seen from Fig. \ref{fig:02}(a)
that for $\lambda=0.08$, the steady states are reached before
$t=40$. Therefore in this case $t_0$ should be greater than $40$ when
evaluating asymptotic $C(t_0,t_0+t)$. The correlation functions
$C(t_0,t_0+t)$ with different $N_s$ and some typical $t_0$ are shown
in Fig. \ref{fig:06}. Here we set $t_0\ge200$ which is much larger
than 40 to ensure that the correlation function is evaluated in steady
state.

In Markovian ($N_s=1$) case, the reduced dynamics only involves one
last time step. Therefore the correlation function obtained in this
case just corresponds the quantum regression theorem results
\cite{carmichael1999-statistical,gardiner2004-quantum,
  breuer2007-the}. It can be seen from Fig. \ref{fig:06} that the
quantum regression theorem misses important non-Markovian effects and
gives invalid results, as already mentioned in Refs.
\cite{alonso2005-multiple,joergensen2019-exploiting}. In Markovian
case [Fig. \ref{fig:06}(a)], correlation function $C(t_0,t_0+t)$ tends
to almost zero no matter what value of $t_0$. With increasing
non-Markovianity, i.e., increasing $N_s$, asymptotic $C(t_0,t_0+t)$
does coherent oscillation with increasing amplitude. This is not
surprising since such phenomenon is similar to asymptotic
$\expval{\sigma_z(t)}$ shown in Fig. \ref{fig:04}.

The behavior of $C(t_0,t_0+t)$ depends on $t_0$, this can be seen
clearly from Fig. \ref{fig:06}(d), where the non-Markovian results are
shown. Rather than always oscillating with large amplitude, the
amplitude varies with different $t_0$. When $t_0=201.6$, the amplitude
can even become very small. This shows that the correlation function
$C(t_0,t_0+t)$ contains much more information than mere observable
$\expval{\sigma_z(t)}$, and the property of correlation function can
not be simply deduced from the behavior of observable.

It can be also seen that with different $t_0$, the coherent
oscillations have phase difference. This means that the value of $t_0$
not only affects the oscillation amplitude but also the oscillation
phase. Due to this phase shift, $t_0=201.6$ does not cause a minimum
oscillation amplitude for $N_s=20$ case. In this case, a $t_0=201.3$
does it (not shown in the figure).

\section{Signal-to-Noise Ratio}
\label{sec:signal-to-noise-ratio}
A typical way to quantify the response to the driving is the
signal-to-noise ratio (SNR)
\cite{mcnamara1989-theory,debnath1989-remarks,loefstedt1994-quantum,gammaitoni1998-stochasticresonance}. QSR
occurs when SNR passes through a maximum as the noise level
increases. The first papers on SR in fact focused on the behavior of
the signal output $\expval{\sigma_z(t)}$, but later the focus shifted
to the SNR both theoretically and experimentally.

The time-averaged asymptotic symmetrized correlation function
$\bar{C}(t)$ coherently oscillates with the driving frequency $\Omega$
when $t$ is large, therefore the power spectrum $\bar{S}(\omega)$
contains a noise background and $\delta$-function peaks at $\Omega$
and its harmonics. The ratio of the coefficient of the fundamental
peak and the value of noise at $\Omega$ is the SNR.

If only considering the fundamental peak, the time-averaged power
spectrum can be described as the superposition of a background noise
power $N(\omega)$ and a $\delta$ signal term for which
\begin{equation}
  \bar{S}(\omega)=G\delta(\omega-\Omega)+N(\omega),
\end{equation}
where $G$ is the strength of the signal. The ratio $G/N(\Omega)$ gives
the SNR. For sufficiently small driving, $N(\omega)$ does not deviate
much from the power spectrum of the undriven system, while for large
driving the effect of the signal on the noise need to be taken into
consideration.

The correlation function $C(t_0,t_0+t)$ can be split into two parts
for which
\begin{equation}
  C(t_0,t_0+t)=C_0(t_0,t_0+t)+C_1(t_0,t_0+t),
\end{equation}
where $C_0$ is the transient part and $C_1$ is the asymptotically
coherent oscillation part. Accordingly the time-averaged asymptotic
correlation function \eqref{eq:time-averaged-C} can be also split into
two parts as
\begin{equation}
  \bar{C}(t)=\bar{C}_0(t)+\bar{C}_1(t).
\end{equation}

\begin{figure}[htbp]
  \centerline{\includegraphics[]{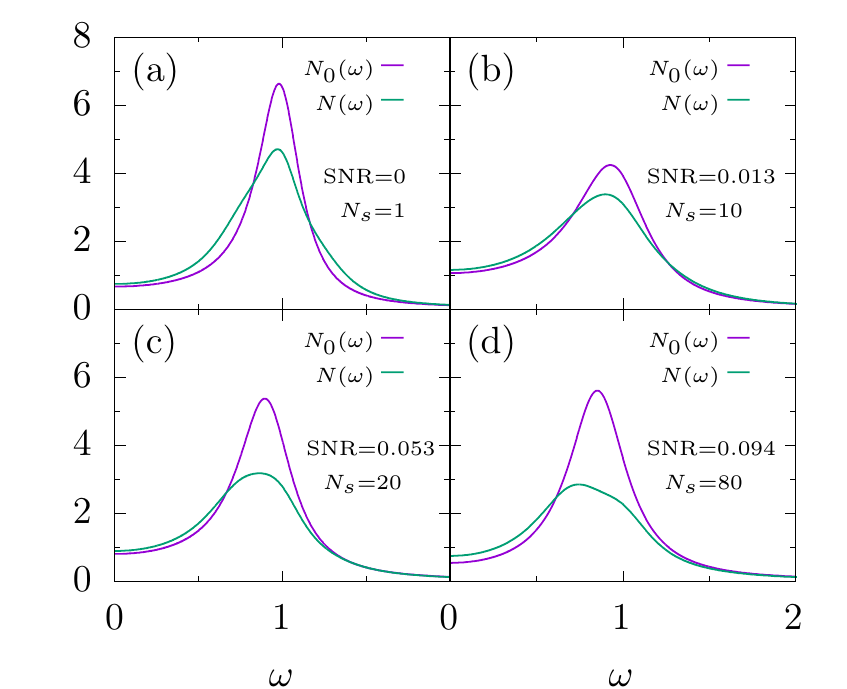}}
  \caption{The background noise power with driving $N(\omega)$ and
    without driving $N_0(\omega)$ for different $N_s$. Here
    $\lambda=0.08$.}
  \label{fig:07}
\end{figure}

The Fourier transform of coherent oscillation part $\bar{C}_1(t)$ just
yields delta peak at $\Omega$ and its harmonics, from which the
strength of signal $G$ is obtained. Here we simply set $G$ as the
amplitude of coherent oscillation $\bar{C}_1(t)$, and background noise
power $N(\omega)$ is the Fourier transform of $\bar{C}_0(t)$. Let
$N_0(\omega)$ be the background noise power without periodic
driving. The background noise powers $N_0(\omega)$ and $N(\omega)$
with different $N_s$ are shown in Fig. \ref{fig:07}.

It should be noted that the shape of $N(\omega)$ presented here is
different to those with Gaussian white noise
\cite{mcnamara1989-theory,gammaitoni1998-stochasticresonance,joshi2008-stochastic}
where $N(\omega)$ is roughly a Lorentzian centered at $\omega=0$.  Our
simulations are in deep quantum regime, the shape of $N(\omega)$ is
roughly an asymmetrical ``Lorentzian'' centered at nonzero
$\omega$. In this sense, we are dealing with the color noise.

From Fig. \ref{fig:07}, it is clear that the driving force alter the
background noise power. In Markovian ($N_s=1$) case, the positions of
peak of $N_0(\omega)$ and $N(\omega)$ are almost the same. But when
$N_s$ increases, the position of peak of $N(\omega)$ starts deviate
from that of $N_0(\omega)$. This is most clear in fully non-Markovian
case, as shown in Fig. \ref{fig:07}(d).

\begin{figure}[htbp]
  \centerline{\includegraphics[]{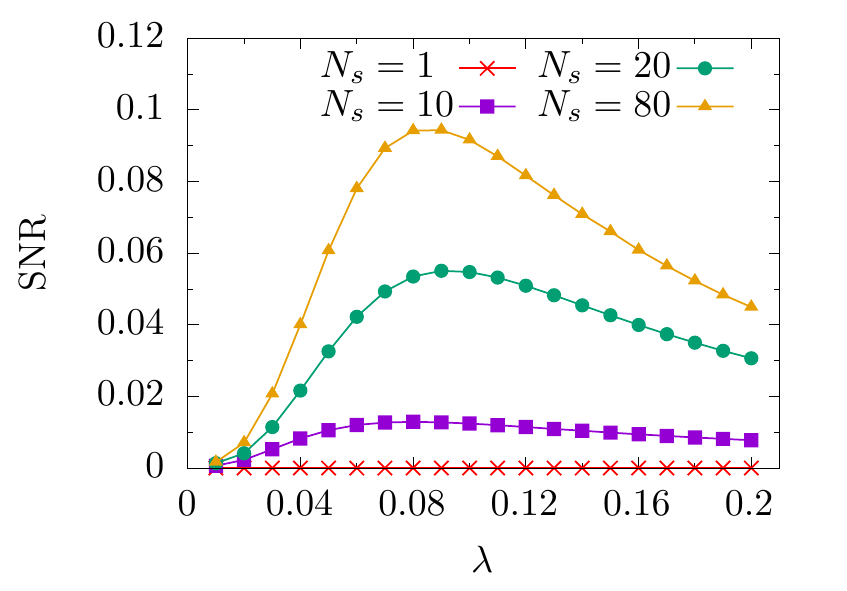}}
  \caption{The signal-to-noise ratio with respect to $\lambda$ for
    different $N_s$.}
\label{fig:08}
\end{figure}

SNR with respect to $\lambda$ for different $N_s$ are shown in
Fig. \ref{fig:08}. It can be seen that, unlike amplitude of
$\expval{\sigma_z(t)}$ shown in Fig. \ref{fig:05}, for weak noise
level $\lambda=0.01$, SNR is almost zero no matter what value of $N_s$
is. Besides, SNR reaches its maximum when $\lambda$ is around 0.08,
which is different from the position of maximum amplitude of
$\expval{\sigma_z(t)}$. This is mostly because the peak position of
background noise power $N(\omega)$ is shifted by non-Markovian
effects.

\section{Conclusions}
\label{sec:conclusions}
In this article, we employ TEMPO algorithm to investigate
non-Markovian effects in QSR. For periodic response signal, the signal
strength is represented by the amplitude of coherent oscillation of
asymptotic $\expval{\sigma_z(t)}$. The QSR is demonstrated by signal
strength as a maximum is presented with increasing noise level.  The
non-Markovianity is controlled by the truncation parameter $N_s$. In
Markovian limit ($N_s=1$), the signal strength is close to zero, which
means the response signal almost vanishes. The signal reappears when
non-Markovianity is included and its strength becomes larger with more
non-Markovianity. This shows the crucial importance of the
non-Markovian effects in QSR.

The correlation function $C(t_0,t_0+t)$ contains much more information
than mere observable $\expval{\sigma_z(t)}$. Unlike
$\expval{\sigma_z(t)}$, whose amplitude simply increases when $N_s$
increases, the amplitude of asymptotic $C(t_0,t_0+t)$ also depends on
the value of $t_0$ in a nontrivial manner. With specific $t_0$, the
amplitude of asymptotic $C(t_0,t_0+t)$ can be close to zero even for
non-Markovian ($N_s=80$) case.

The time-averaged correlation function $\bar{C}(t)$ is obtained by
averaging correlation function $C(t_0,t_0+t)$ with respect to $t_0$
over a period. This $\bar{C}(t)$ can be split into a transient part
$\bar{C}_0(t)$ and a coherent oscillation part $\bar{C}_1(t)$. The
Fourier transform of $\bar{C}_0(t)$ gives the background noise power
$N(\omega)$, and the amplitude of $\bar{C}_1(t)$ gives the signal
strength. The $N(\omega)$ obtained in our model differs from that from
white noise approximation, which indicates that effects of environment
in deep quantum regime can be viewed as color noise. When comparing
$N(\omega)$ to the background noise power without driving
$N_0(\omega)$, it is found that their peaks are at almost the same
position in Markovian limit. When non-Markovianity is included, the
position of their peaks deviate from each other, and this is most
clearly seen in fully non-Markovian case. This makes the maximum of
SNR and $\expval{\sigma_z(t)}$ appear at different noise level.

\textit{Acknowledgments}. This work is supported by the NSFC Grant
No. 12104328.


\begin{thebibliography}{41}%
\makeatletter
\providecommand \@ifxundefined [1]{%
 \@ifx{#1\undefined}
}%
\providecommand \@ifnum [1]{%
 \ifnum #1\expandafter \@firstoftwo
 \else \expandafter \@secondoftwo
 \fi
}%
\providecommand \@ifx [1]{%
 \ifx #1\expandafter \@firstoftwo
 \else \expandafter \@secondoftwo
 \fi
}%
\providecommand \natexlab [1]{#1}%
\providecommand \enquote  [1]{``#1''}%
\providecommand \bibnamefont  [1]{#1}%
\providecommand \bibfnamefont [1]{#1}%
\providecommand \citenamefont [1]{#1}%
\providecommand \href@noop [0]{\@secondoftwo}%
\providecommand \href [0]{\begingroup \@sanitize@url \@href}%
\providecommand \@href[1]{\@@startlink{#1}\@@href}%
\providecommand \@@href[1]{\endgroup#1\@@endlink}%
\providecommand \@sanitize@url [0]{\catcode `\\12\catcode `\$12\catcode
  `\&12\catcode `\#12\catcode `\^12\catcode `\_12\catcode `\%12\relax}%
\providecommand \@@startlink[1]{}%
\providecommand \@@endlink[0]{}%
\providecommand \url  [0]{\begingroup\@sanitize@url \@url }%
\providecommand \@url [1]{\endgroup\@href {#1}{\urlprefix }}%
\providecommand \urlprefix  [0]{URL }%
\providecommand \Eprint [0]{\href }%
\providecommand \doibase [0]{http://dx.doi.org/}%
\providecommand \selectlanguage [0]{\@gobble}%
\providecommand \bibinfo  [0]{\@secondoftwo}%
\providecommand \bibfield  [0]{\@secondoftwo}%
\providecommand \translation [1]{[#1]}%
\providecommand \BibitemOpen [0]{}%
\providecommand \bibitemStop [0]{}%
\providecommand \bibitemNoStop [0]{.\EOS\space}%
\providecommand \EOS [0]{\spacefactor3000\relax}%
\providecommand \BibitemShut  [1]{\csname bibitem#1\endcsname}%
\let\auto@bib@innerbib\@empty
\bibitem [{\citenamefont {Benzi}\ \emph {et~al.}(1981)\citenamefont {Benzi},
  \citenamefont {Sutera},\ and\ \citenamefont {Vulpiani}}]{benzi1981-the}%
  \BibitemOpen
  \bibfield  {author} {\bibinfo {author} {\bibfnamefont {R.}~\bibnamefont
  {Benzi}}, \bibinfo {author} {\bibfnamefont {A.}~\bibnamefont {Sutera}}, \
  and\ \bibinfo {author} {\bibfnamefont {A.}~\bibnamefont {Vulpiani}},\ }\href
  {\doibase 10.1088/0305-4470/14/11/006} {\bibfield  {journal} {\bibinfo
  {journal} {Journal of Physics A: Mathematical and General}\ }\textbf
  {\bibinfo {volume} {14}},\ \bibinfo {pages} {L453} (\bibinfo {year}
  {1981})}\BibitemShut {NoStop}%
\bibitem [{\citenamefont {Fauve}\ and\ \citenamefont
  {Heslot}(1983)}]{fauve1983-stochastic}%
  \BibitemOpen
  \bibfield  {author} {\bibinfo {author} {\bibfnamefont {S.}~\bibnamefont
  {Fauve}}\ and\ \bibinfo {author} {\bibfnamefont {F.}~\bibnamefont {Heslot}},\
  }\href {\doibase 10.1016/0375-9601(83)90086-5} {\bibfield  {journal}
  {\bibinfo  {journal} {Physics Letters A}\ }\textbf {\bibinfo {volume} {97}},\
  \bibinfo {pages} {5} (\bibinfo {year} {1983})}\BibitemShut {NoStop}%
\bibitem [{\citenamefont {McNamara}\ \emph {et~al.}(1988)\citenamefont
  {McNamara}, \citenamefont {Wiesenfeld},\ and\ \citenamefont
  {Roy}}]{mcnamara1988-observation}%
  \BibitemOpen
  \bibfield  {author} {\bibinfo {author} {\bibfnamefont {B.}~\bibnamefont
  {McNamara}}, \bibinfo {author} {\bibfnamefont {K.}~\bibnamefont
  {Wiesenfeld}}, \ and\ \bibinfo {author} {\bibfnamefont {R.}~\bibnamefont
  {Roy}},\ }\href {\doibase 10.1103/physrevlett.60.2626} {\bibfield  {journal}
  {\bibinfo  {journal} {Physical Review Letters}\ }\textbf {\bibinfo {volume}
  {60}},\ \bibinfo {pages} {2626} (\bibinfo {year} {1988})}\BibitemShut
  {NoStop}%
\bibitem [{\citenamefont {L{\"o}fstedt}\ and\ \citenamefont
  {Coppersmith}(1994)}]{loefstedt1994-quantum}%
  \BibitemOpen
  \bibfield  {author} {\bibinfo {author} {\bibfnamefont {R.}~\bibnamefont
  {L{\"o}fstedt}}\ and\ \bibinfo {author} {\bibfnamefont {S.~N.}\ \bibnamefont
  {Coppersmith}},\ }\href {\doibase 10.1103/physrevlett.72.1947} {\bibfield
  {journal} {\bibinfo  {journal} {Physical Review Letters}\ }\textbf {\bibinfo
  {volume} {72}},\ \bibinfo {pages} {1947} (\bibinfo {year}
  {1994})}\BibitemShut {NoStop}%
\bibitem [{\citenamefont {Gammaitoni}\ \emph {et~al.}(1999)\citenamefont
  {Gammaitoni}, \citenamefont {L{\"o}cher}, \citenamefont {Bulsara},
  \citenamefont {H{\"a}nggi}, \citenamefont {Neff}, \citenamefont {Wiesenfeld},
  \citenamefont {Ditto},\ and\ \citenamefont
  {Inchiosa}}]{gammaitoni1999-controlling}%
  \BibitemOpen
  \bibfield  {author} {\bibinfo {author} {\bibfnamefont {L.}~\bibnamefont
  {Gammaitoni}}, \bibinfo {author} {\bibfnamefont {M.}~\bibnamefont
  {L{\"o}cher}}, \bibinfo {author} {\bibfnamefont {A.}~\bibnamefont {Bulsara}},
  \bibinfo {author} {\bibfnamefont {P.}~\bibnamefont {H{\"a}nggi}}, \bibinfo
  {author} {\bibfnamefont {J.}~\bibnamefont {Neff}}, \bibinfo {author}
  {\bibfnamefont {K.}~\bibnamefont {Wiesenfeld}}, \bibinfo {author}
  {\bibfnamefont {W.}~\bibnamefont {Ditto}}, \ and\ \bibinfo {author}
  {\bibfnamefont {M.~E.}\ \bibnamefont {Inchiosa}},\ }\href {\doibase
  10.1103/physrevlett.82.4574} {\bibfield  {journal} {\bibinfo  {journal}
  {Physical Review Letters}\ }\textbf {\bibinfo {volume} {82}},\ \bibinfo
  {pages} {4574} (\bibinfo {year} {1999})}\BibitemShut {NoStop}%
\bibitem [{\citenamefont {Wagner}\ \emph {et~al.}(2019)\citenamefont {Wagner},
  \citenamefont {Talkner}, \citenamefont {Bayer}, \citenamefont {Rugeramigabo},
  \citenamefont {H{\"a}nggi},\ and\ \citenamefont {Haug}}]{wagner2019-quantum}%
  \BibitemOpen
  \bibfield  {author} {\bibinfo {author} {\bibfnamefont {T.}~\bibnamefont
  {Wagner}}, \bibinfo {author} {\bibfnamefont {P.}~\bibnamefont {Talkner}},
  \bibinfo {author} {\bibfnamefont {J.~C.}\ \bibnamefont {Bayer}}, \bibinfo
  {author} {\bibfnamefont {E.~P.}\ \bibnamefont {Rugeramigabo}}, \bibinfo
  {author} {\bibfnamefont {P.}~\bibnamefont {H{\"a}nggi}}, \ and\ \bibinfo
  {author} {\bibfnamefont {R.~J.}\ \bibnamefont {Haug}},\ }\href {\doibase
  10.1038/s41567-018-0412-5} {\bibfield  {journal} {\bibinfo  {journal} {Nature
  Physics}\ }\textbf {\bibinfo {volume} {15}},\ \bibinfo {pages} {330}
  (\bibinfo {year} {2019})}\BibitemShut {NoStop}%
\bibitem [{\citenamefont {Hussein}\ \emph {et~al.}(2020)\citenamefont
  {Hussein}, \citenamefont {Kohler}, \citenamefont {Bayer}, \citenamefont
  {Wagner},\ and\ \citenamefont {Haug}}]{hussein2020-spectral}%
  \BibitemOpen
  \bibfield  {author} {\bibinfo {author} {\bibfnamefont {R.}~\bibnamefont
  {Hussein}}, \bibinfo {author} {\bibfnamefont {S.}~\bibnamefont {Kohler}},
  \bibinfo {author} {\bibfnamefont {J.~C.}\ \bibnamefont {Bayer}}, \bibinfo
  {author} {\bibfnamefont {T.}~\bibnamefont {Wagner}}, \ and\ \bibinfo {author}
  {\bibfnamefont {R.~J.}\ \bibnamefont {Haug}},\ }\href {\doibase
  10.1103/physrevlett.125.206801} {\bibfield  {journal} {\bibinfo  {journal}
  {Physical Review Letters}\ }\textbf {\bibinfo {volume} {125}},\ \bibinfo
  {pages} {206801} (\bibinfo {year} {2020})}\BibitemShut {NoStop}%
\bibitem [{\citenamefont {H{\"a}nze}\ \emph {et~al.}(2021)\citenamefont
  {H{\"a}nze}, \citenamefont {McMurtrie}, \citenamefont {Baumann},
  \citenamefont {Malavolti}, \citenamefont {Coppersmith},\ and\ \citenamefont
  {Loth}}]{haenze2021-quantum}%
  \BibitemOpen
  \bibfield  {author} {\bibinfo {author} {\bibfnamefont {M.}~\bibnamefont
  {H{\"a}nze}}, \bibinfo {author} {\bibfnamefont {G.}~\bibnamefont
  {McMurtrie}}, \bibinfo {author} {\bibfnamefont {S.}~\bibnamefont {Baumann}},
  \bibinfo {author} {\bibfnamefont {L.}~\bibnamefont {Malavolti}}, \bibinfo
  {author} {\bibfnamefont {S.~N.}\ \bibnamefont {Coppersmith}}, \ and\ \bibinfo
  {author} {\bibfnamefont {S.}~\bibnamefont {Loth}},\ }\href {\doibase
  10.1126/sciadv.abg2616} {\bibfield  {journal} {\bibinfo  {journal} {Science
  Advances}\ }\textbf {\bibinfo {volume} {7}},\ \bibinfo {pages} {eabg2616}
  (\bibinfo {year} {2021})}\BibitemShut {NoStop}%
\bibitem [{\citenamefont {McNamara}\ and\ \citenamefont
  {Wiesenfeld}(1989)}]{mcnamara1989-theory}%
  \BibitemOpen
  \bibfield  {author} {\bibinfo {author} {\bibfnamefont {B.}~\bibnamefont
  {McNamara}}\ and\ \bibinfo {author} {\bibfnamefont {K.}~\bibnamefont
  {Wiesenfeld}},\ }\href {\doibase 10.1103/physreva.39.4854} {\bibfield
  {journal} {\bibinfo  {journal} {Physical Review A}\ }\textbf {\bibinfo
  {volume} {39}},\ \bibinfo {pages} {4854} (\bibinfo {year}
  {1989})}\BibitemShut {NoStop}%
\bibitem [{\citenamefont {Jung}(1993)}]{jung1993-periodically}%
  \BibitemOpen
  \bibfield  {author} {\bibinfo {author} {\bibfnamefont {P.}~\bibnamefont
  {Jung}},\ }\href {\doibase 10.1016/0370-1573(93)90022-6} {\bibfield
  {journal} {\bibinfo  {journal} {Physics Reports}\ }\textbf {\bibinfo {volume}
  {234}},\ \bibinfo {pages} {175} (\bibinfo {year} {1993})}\BibitemShut
  {NoStop}%
\bibitem [{\citenamefont {Grifoni}\ and\ \citenamefont
  {H\"anggi}(1998)}]{grifoni1998-driven}%
  \BibitemOpen
  \bibfield  {author} {\bibinfo {author} {\bibfnamefont {M.}~\bibnamefont
  {Grifoni}}\ and\ \bibinfo {author} {\bibfnamefont {P.}~\bibnamefont
  {H\"anggi}},\ }\href {\doibase 10.1016/S0370-1573(98)00022-2} {\bibfield
  {journal} {\bibinfo  {journal} {Physics Reports}\ }\textbf {\bibinfo {volume}
  {304}},\ \bibinfo {pages} {229} (\bibinfo {year} {1998})}\BibitemShut
  {NoStop}%
\bibitem [{\citenamefont {Gammaitoni}\ \emph {et~al.}(1998)\citenamefont
  {Gammaitoni}, \citenamefont {H{\"a}nggi}, \citenamefont {Jung},\ and\
  \citenamefont {Marchesoni}}]{gammaitoni1998-stochasticresonance}%
  \BibitemOpen
  \bibfield  {author} {\bibinfo {author} {\bibfnamefont {L.}~\bibnamefont
  {Gammaitoni}}, \bibinfo {author} {\bibfnamefont {P.}~\bibnamefont
  {H{\"a}nggi}}, \bibinfo {author} {\bibfnamefont {P.}~\bibnamefont {Jung}}, \
  and\ \bibinfo {author} {\bibfnamefont {F.}~\bibnamefont {Marchesoni}},\
  }\href {\doibase 10.1103/revmodphys.70.223} {\bibfield  {journal} {\bibinfo
  {journal} {Reviews of Modern Physics}\ }\textbf {\bibinfo {volume} {70}},\
  \bibinfo {pages} {223} (\bibinfo {year} {1998})}\BibitemShut {NoStop}%
\bibitem [{\citenamefont {Grifoni}\ \emph {et~al.}(1996)\citenamefont
  {Grifoni}, \citenamefont {Hartmann}, \citenamefont {Berchtold},\ and\
  \citenamefont {H{\"a}nggi}}]{grifoni1996-quantum}%
  \BibitemOpen
  \bibfield  {author} {\bibinfo {author} {\bibfnamefont {M.}~\bibnamefont
  {Grifoni}}, \bibinfo {author} {\bibfnamefont {L.}~\bibnamefont {Hartmann}},
  \bibinfo {author} {\bibfnamefont {S.}~\bibnamefont {Berchtold}}, \ and\
  \bibinfo {author} {\bibfnamefont {P.}~\bibnamefont {H{\"a}nggi}},\ }\href
  {\doibase 10.1103/physreve.53.5890} {\bibfield  {journal} {\bibinfo
  {journal} {Physical Review E}\ }\textbf {\bibinfo {volume} {53}},\ \bibinfo
  {pages} {5890} (\bibinfo {year} {1996})}\BibitemShut {NoStop}%
\bibitem [{\citenamefont {Grifoni}\ and\ \citenamefont
  {H{\"a}nggi}(1996)}]{grifoni1996-coherent}%
  \BibitemOpen
  \bibfield  {author} {\bibinfo {author} {\bibfnamefont {M.}~\bibnamefont
  {Grifoni}}\ and\ \bibinfo {author} {\bibfnamefont {P.}~\bibnamefont
  {H{\"a}nggi}},\ }\href {\doibase 10.1103/physrevlett.76.1611} {\bibfield
  {journal} {\bibinfo  {journal} {Physical Review Letters}\ }\textbf {\bibinfo
  {volume} {76}},\ \bibinfo {pages} {1611} (\bibinfo {year}
  {1996})}\BibitemShut {NoStop}%
\bibitem [{\citenamefont {Makarov}\ and\ \citenamefont
  {Makri}(1995{\natexlab{a}})}]{makarov1995-stochastic}%
  \BibitemOpen
  \bibfield  {author} {\bibinfo {author} {\bibfnamefont {D.~E.}\ \bibnamefont
  {Makarov}}\ and\ \bibinfo {author} {\bibfnamefont {N.}~\bibnamefont
  {Makri}},\ }\href {\doibase 10.1103/physrevb.52.r2257} {\bibfield  {journal}
  {\bibinfo  {journal} {Physical Review B}\ }\textbf {\bibinfo {volume} {52}},\
  \bibinfo {pages} {R2257} (\bibinfo {year} {1995}{\natexlab{a}})}\BibitemShut
  {NoStop}%
\bibitem [{\citenamefont {Makarov}\ and\ \citenamefont
  {Makri}(1995{\natexlab{b}})}]{makarov1995-control}%
  \BibitemOpen
  \bibfield  {author} {\bibinfo {author} {\bibfnamefont {D.~E.}\ \bibnamefont
  {Makarov}}\ and\ \bibinfo {author} {\bibfnamefont {N.}~\bibnamefont
  {Makri}},\ }\href {\doibase 10.1103/physreve.52.5863} {\bibfield  {journal}
  {\bibinfo  {journal} {Physical Review E}\ }\textbf {\bibinfo {volume} {52}},\
  \bibinfo {pages} {5863} (\bibinfo {year} {1995}{\natexlab{b}})}\BibitemShut
  {NoStop}%
\bibitem [{\citenamefont {Joshi}(2008)}]{joshi2008-stochastic}%
  \BibitemOpen
  \bibfield  {author} {\bibinfo {author} {\bibfnamefont {A.}~\bibnamefont
  {Joshi}},\ }\href {\doibase 10.1103/physreve.77.020104} {\bibfield  {journal}
  {\bibinfo  {journal} {Physical Review E}\ }\textbf {\bibinfo {volume} {77}},\
  \bibinfo {pages} {020104} (\bibinfo {year} {2008})}\BibitemShut {NoStop}%
\bibitem [{\citenamefont {Strathearn}\ \emph {et~al.}(2018)\citenamefont
  {Strathearn}, \citenamefont {Kirton}, \citenamefont {Kilda}, \citenamefont
  {Keeling},\ and\ \citenamefont {Lovett}}]{strathearn2018-efficient}%
  \BibitemOpen
  \bibfield  {author} {\bibinfo {author} {\bibfnamefont {A.}~\bibnamefont
  {Strathearn}}, \bibinfo {author} {\bibfnamefont {P.}~\bibnamefont {Kirton}},
  \bibinfo {author} {\bibfnamefont {D.}~\bibnamefont {Kilda}}, \bibinfo
  {author} {\bibfnamefont {J.}~\bibnamefont {Keeling}}, \ and\ \bibinfo
  {author} {\bibfnamefont {B.~W.}\ \bibnamefont {Lovett}},\ }\href {\doibase
  10.1038/s41467-018-05617-3} {\bibfield  {journal} {\bibinfo  {journal}
  {Nature Communications}\ }\textbf {\bibinfo {volume} {9}},\ \bibinfo {pages}
  {3322} (\bibinfo {year} {2018})}\BibitemShut {NoStop}%
\bibitem [{\citenamefont {Schollw{\"o}ck}(2011)}]{schollwoeck2011-density}%
  \BibitemOpen
  \bibfield  {author} {\bibinfo {author} {\bibfnamefont {U.}~\bibnamefont
  {Schollw{\"o}ck}},\ }\href {\doibase 10.1016/j.aop.2010.09.012} {\bibfield
  {journal} {\bibinfo  {journal} {Annals of Physics}\ }\textbf {\bibinfo
  {volume} {326}},\ \bibinfo {pages} {96} (\bibinfo {year} {2011})}\BibitemShut
  {NoStop}%
\bibitem [{\citenamefont {Or{\'u}s}(2014)}]{orus2014-practical}%
  \BibitemOpen
  \bibfield  {author} {\bibinfo {author} {\bibfnamefont {R.}~\bibnamefont
  {Or{\'u}s}},\ }\href {\doibase 10.1016/j.aop.2014.06.013} {\bibfield
  {journal} {\bibinfo  {journal} {Annals of Physics}\ }\textbf {\bibinfo
  {volume} {349}},\ \bibinfo {pages} {117} (\bibinfo {year}
  {2014})}\BibitemShut {NoStop}%
\bibitem [{\citenamefont {J{\o}rgensen}\ and\ \citenamefont
  {Pollock}(2019)}]{joergensen2019-exploiting}%
  \BibitemOpen
  \bibfield  {author} {\bibinfo {author} {\bibfnamefont {M.~R.}\ \bibnamefont
  {J{\o}rgensen}}\ and\ \bibinfo {author} {\bibfnamefont {F.~A.}\ \bibnamefont
  {Pollock}},\ }\href {\doibase 10.1103/physrevlett.123.240602} {\bibfield
  {journal} {\bibinfo  {journal} {Physical Review Letters}\ }\textbf {\bibinfo
  {volume} {123}},\ \bibinfo {pages} {240602} (\bibinfo {year}
  {2019})}\BibitemShut {NoStop}%
\bibitem [{\citenamefont {Fux}\ \emph {et~al.}(2021)\citenamefont {Fux},
  \citenamefont {Butler}, \citenamefont {Eastham}, \citenamefont {Lovett},\
  and\ \citenamefont {Keeling}}]{fux2021-efficient}%
  \BibitemOpen
  \bibfield  {author} {\bibinfo {author} {\bibfnamefont {G.~E.}\ \bibnamefont
  {Fux}}, \bibinfo {author} {\bibfnamefont {E.~P.}\ \bibnamefont {Butler}},
  \bibinfo {author} {\bibfnamefont {P.~R.}\ \bibnamefont {Eastham}}, \bibinfo
  {author} {\bibfnamefont {B.~W.}\ \bibnamefont {Lovett}}, \ and\ \bibinfo
  {author} {\bibfnamefont {J.}~\bibnamefont {Keeling}},\ }\href {\doibase
  10.1103/physrevlett.126.200401} {\bibfield  {journal} {\bibinfo  {journal}
  {Physical Review Letters}\ }\textbf {\bibinfo {volume} {126}},\ \bibinfo
  {pages} {200401} (\bibinfo {year} {2021})}\BibitemShut {NoStop}%
\bibitem [{\citenamefont {Caldeira}\ and\ \citenamefont
  {Leggett}(1983{\natexlab{a}})}]{caldeira1983-path}%
  \BibitemOpen
  \bibfield  {author} {\bibinfo {author} {\bibfnamefont {A.~O.}\ \bibnamefont
  {Caldeira}}\ and\ \bibinfo {author} {\bibfnamefont {A.~J.}\ \bibnamefont
  {Leggett}},\ }\href {\doibase 10.1016/0378-4371(83)90013-4} {\bibfield
  {journal} {\bibinfo  {journal} {Physica A}\ }\textbf {\bibinfo {volume}
  {121}},\ \bibinfo {pages} {587} (\bibinfo {year}
  {1983}{\natexlab{a}})}\BibitemShut {NoStop}%
\bibitem [{\citenamefont {Caldeira}\ and\ \citenamefont
  {Leggett}(1983{\natexlab{b}})}]{caldeira1983-quantum}%
  \BibitemOpen
  \bibfield  {author} {\bibinfo {author} {\bibfnamefont {A.}~\bibnamefont
  {Caldeira}}\ and\ \bibinfo {author} {\bibfnamefont {A.}~\bibnamefont
  {Leggett}},\ }\href {\doibase 10.1016/0003-4916(83)90202-6} {\bibfield
  {journal} {\bibinfo  {journal} {Annals of Physics}\ }\textbf {\bibinfo
  {volume} {149}},\ \bibinfo {pages} {374} (\bibinfo {year}
  {1983}{\natexlab{b}})}\BibitemShut {NoStop}%
\bibitem [{\citenamefont {Leggett}\ \emph {et~al.}(1987)\citenamefont
  {Leggett}, \citenamefont {Chakravarty}, \citenamefont {Dorsey}, \citenamefont
  {Fisher}, \citenamefont {Garg},\ and\ \citenamefont
  {Zwerger}}]{leggett1987-dynamics}%
  \BibitemOpen
  \bibfield  {author} {\bibinfo {author} {\bibfnamefont {A.~J.}\ \bibnamefont
  {Leggett}}, \bibinfo {author} {\bibfnamefont {S.}~\bibnamefont
  {Chakravarty}}, \bibinfo {author} {\bibfnamefont {A.~T.}\ \bibnamefont
  {Dorsey}}, \bibinfo {author} {\bibfnamefont {M.~P.~A.}\ \bibnamefont
  {Fisher}}, \bibinfo {author} {\bibfnamefont {A.}~\bibnamefont {Garg}}, \ and\
  \bibinfo {author} {\bibfnamefont {W.}~\bibnamefont {Zwerger}},\ }\href
  {\doibase 10.1103/revmodphys.59.1} {\bibfield  {journal} {\bibinfo  {journal}
  {Reviews of Modern Physics}\ }\textbf {\bibinfo {volume} {59}},\ \bibinfo
  {pages} {1} (\bibinfo {year} {1987})}\BibitemShut {NoStop}%
\bibitem [{\citenamefont {Weiss}(1993)}]{weiss1993-quantum}%
  \BibitemOpen
  \bibfield  {author} {\bibinfo {author} {\bibfnamefont {U.}~\bibnamefont
  {Weiss}},\ }\href {\doibase 10.1142/8334} {\emph {\bibinfo {title} {Quantum
  Dissipative Systems}}}\ (\bibinfo  {publisher} {World Scientific},\ \bibinfo
  {address} {Singapore},\ \bibinfo {year} {1993})\BibitemShut {NoStop}%
\bibitem [{\citenamefont {Feynman}\ and\ \citenamefont
  {Vernon}(1963)}]{feynman1963-the}%
  \BibitemOpen
  \bibfield  {author} {\bibinfo {author} {\bibfnamefont {R.~P.}\ \bibnamefont
  {Feynman}}\ and\ \bibinfo {author} {\bibfnamefont {F.~L.}\ \bibnamefont
  {Vernon}},\ }\href {\doibase 10.1016/0003-4916(63)90068-X} {\bibfield
  {journal} {\bibinfo  {journal} {Annals of Physics}\ }\textbf {\bibinfo
  {volume} {24}},\ \bibinfo {pages} {118} (\bibinfo {year} {1963})}\BibitemShut
  {NoStop}%
\bibitem [{\citenamefont {Feynman}\ and\ \citenamefont
  {Hibbs}(1965)}]{feynman1965-quantum}%
  \BibitemOpen
  \bibfield  {author} {\bibinfo {author} {\bibfnamefont {R.~P.}\ \bibnamefont
  {Feynman}}\ and\ \bibinfo {author} {\bibfnamefont {A.~R.}\ \bibnamefont
  {Hibbs}},\ }\href@noop {} {\emph {\bibinfo {title} {Quantum Mechanics and
  Path Integrals}}}\ (\bibinfo  {publisher} {Mc Graw-Hill, New York},\ \bibinfo
  {year} {1965})\BibitemShut {NoStop}%
\bibitem [{\citenamefont {Dattani}\ \emph {et~al.}(2012)\citenamefont
  {Dattani}, \citenamefont {Pollock},\ and\ \citenamefont
  {Wilkins}}]{dattani2012-analytic}%
  \BibitemOpen
  \bibfield  {author} {\bibinfo {author} {\bibfnamefont {N.~S.}\ \bibnamefont
  {Dattani}}, \bibinfo {author} {\bibfnamefont {F.~A.}\ \bibnamefont
  {Pollock}}, \ and\ \bibinfo {author} {\bibfnamefont {D.~M.}\ \bibnamefont
  {Wilkins}},\ }\href
  {http://www.naturalspublishing.com/Article.asp?ArtcID=407} {\bibfield
  {journal} {\bibinfo  {journal} {Quantum Physics Letters}\ }\textbf {\bibinfo
  {volume} {1}},\ \bibinfo {pages} {35} (\bibinfo {year} {2012})}\BibitemShut
  {NoStop}%
\bibitem [{\citenamefont {Makarov}\ and\ \citenamefont
  {Makri}(1993)}]{makarov1993-tunneling}%
  \BibitemOpen
  \bibfield  {author} {\bibinfo {author} {\bibfnamefont {D.~E.}\ \bibnamefont
  {Makarov}}\ and\ \bibinfo {author} {\bibfnamefont {N.}~\bibnamefont
  {Makri}},\ }\href {\doibase 10.1103/physreva.48.3626} {\bibfield  {journal}
  {\bibinfo  {journal} {Physical Review A}\ }\textbf {\bibinfo {volume} {48}},\
  \bibinfo {pages} {3626} (\bibinfo {year} {1993})}\BibitemShut {NoStop}%
\bibitem [{\citenamefont {Makarov}\ and\ \citenamefont
  {Makri}(1994)}]{makarov1994-path}%
  \BibitemOpen
  \bibfield  {author} {\bibinfo {author} {\bibfnamefont {D.~E.}\ \bibnamefont
  {Makarov}}\ and\ \bibinfo {author} {\bibfnamefont {N.}~\bibnamefont
  {Makri}},\ }\href {\doibase 10.1016/0009-2614(94)00275-4} {\bibfield
  {journal} {\bibinfo  {journal} {Chemical Physics Letters}\ }\textbf {\bibinfo
  {volume} {221}},\ \bibinfo {pages} {482} (\bibinfo {year}
  {1994})}\BibitemShut {NoStop}%
\bibitem [{\citenamefont {Makri}(1995)}]{makri1995-numerical}%
  \BibitemOpen
  \bibfield  {author} {\bibinfo {author} {\bibfnamefont {N.}~\bibnamefont
  {Makri}},\ }\href {\doibase 10.1063/1.531046} {\bibfield  {journal} {\bibinfo
   {journal} {Journal of Mathematical Physics}\ }\textbf {\bibinfo {volume}
  {36}},\ \bibinfo {pages} {2430} (\bibinfo {year} {1995})}\BibitemShut
  {NoStop}%
\bibitem [{\citenamefont {Oshiyama}\ \emph {et~al.}(2020)\citenamefont
  {Oshiyama}, \citenamefont {Shibata},\ and\ \citenamefont
  {Suzuki}}]{oshiyama2020-kibble}%
  \BibitemOpen
  \bibfield  {author} {\bibinfo {author} {\bibfnamefont {H.}~\bibnamefont
  {Oshiyama}}, \bibinfo {author} {\bibfnamefont {N.}~\bibnamefont {Shibata}}, \
  and\ \bibinfo {author} {\bibfnamefont {S.}~\bibnamefont {Suzuki}},\ }\href
  {\doibase 10.7566/jpsj.89.104002} {\bibfield  {journal} {\bibinfo  {journal}
  {Journal of the Physical Society of Japan}\ }\textbf {\bibinfo {volume}
  {89}},\ \bibinfo {pages} {104002} (\bibinfo {year} {2020})}\BibitemShut
  {NoStop}%
\bibitem [{\citenamefont {Oshiyama}\ \emph {et~al.}(2022)\citenamefont
  {Oshiyama}, \citenamefont {Suzuki},\ and\ \citenamefont
  {Shibata}}]{oshiyama2022-classical}%
  \BibitemOpen
  \bibfield  {author} {\bibinfo {author} {\bibfnamefont {H.}~\bibnamefont
  {Oshiyama}}, \bibinfo {author} {\bibfnamefont {S.}~\bibnamefont {Suzuki}}, \
  and\ \bibinfo {author} {\bibfnamefont {N.}~\bibnamefont {Shibata}},\ }\href
  {\doibase 10.1103/physrevlett.128.170502} {\bibfield  {journal} {\bibinfo
  {journal} {Physical Review Letters}\ }\textbf {\bibinfo {volume} {128}},\
  \bibinfo {pages} {170502} (\bibinfo {year} {2022})}\BibitemShut {NoStop}%
\bibitem [{\citenamefont {Nalbach}\ \emph {et~al.}(2011)\citenamefont
  {Nalbach}, \citenamefont {Ishizaki}, \citenamefont {Fleming},\ and\
  \citenamefont {Thorwart}}]{nalbach2011-iterative}%
  \BibitemOpen
  \bibfield  {author} {\bibinfo {author} {\bibfnamefont {P.}~\bibnamefont
  {Nalbach}}, \bibinfo {author} {\bibfnamefont {A.}~\bibnamefont {Ishizaki}},
  \bibinfo {author} {\bibfnamefont {G.~R.}\ \bibnamefont {Fleming}}, \ and\
  \bibinfo {author} {\bibfnamefont {M.}~\bibnamefont {Thorwart}},\ }\href
  {\doibase 10.1088/1367-2630/13/6/063040} {\bibfield  {journal} {\bibinfo
  {journal} {New Journal of Physics}\ }\textbf {\bibinfo {volume} {13}},\
  \bibinfo {pages} {063040} (\bibinfo {year} {2011})}\BibitemShut {NoStop}%
\bibitem [{\citenamefont {Thorwart}\ \emph {et~al.}(2005)\citenamefont
  {Thorwart}, \citenamefont {Eckel},\ and\ \citenamefont
  {Mucciolo}}]{thorwart2005-non}%
  \BibitemOpen
  \bibfield  {author} {\bibinfo {author} {\bibfnamefont {M.}~\bibnamefont
  {Thorwart}}, \bibinfo {author} {\bibfnamefont {J.}~\bibnamefont {Eckel}}, \
  and\ \bibinfo {author} {\bibfnamefont {E.~R.}\ \bibnamefont {Mucciolo}},\
  }\href {\doibase 10.1103/physrevb.72.235320} {\bibfield  {journal} {\bibinfo
  {journal} {Physical Review B}\ }\textbf {\bibinfo {volume} {72}},\ \bibinfo
  {pages} {235320} (\bibinfo {year} {2005})}\BibitemShut {NoStop}%
\bibitem [{\citenamefont {Carmichael}(1999)}]{carmichael1999-statistical}%
  \BibitemOpen
  \bibfield  {author} {\bibinfo {author} {\bibfnamefont {H.~J.}\ \bibnamefont
  {Carmichael}},\ }\href@noop {} {\emph {\bibinfo {title} {Statistical Methods
  in Quantum Optics 1: Master Equations and Fokker-Planck Equations}}}\
  (\bibinfo  {publisher} {Springer-Verlag Berlin},\ \bibinfo {address}
  {Berlin},\ \bibinfo {year} {1999})\BibitemShut {NoStop}%
\bibitem [{\citenamefont {Gardiner}\ and\ \citenamefont
  {Zoller}(2004)}]{gardiner2004-quantum}%
  \BibitemOpen
  \bibfield  {author} {\bibinfo {author} {\bibfnamefont {C.~W.}\ \bibnamefont
  {Gardiner}}\ and\ \bibinfo {author} {\bibfnamefont {P.}~\bibnamefont
  {Zoller}},\ }\href@noop {} {\emph {\bibinfo {title} {Quantum Noise: A
  Handbook of Markovian and Non-Markovian Quantum Stochastic Methods with
  Applications to Quantum Optics}}}\ (\bibinfo  {publisher} {Springer-Verlag},\
  \bibinfo {address} {Berlin},\ \bibinfo {year} {2004})\BibitemShut {NoStop}%
\bibitem [{\citenamefont {Breuer}(2007)}]{breuer2007-the}%
  \BibitemOpen
  \bibfield  {author} {\bibinfo {author} {\bibfnamefont {H.-P.}\ \bibnamefont
  {Breuer}},\ }\href@noop {} {\emph {\bibinfo {title} {The Theory of Open
  Quantum Systems}}}\ (\bibinfo  {publisher} {Oxford University Press},\
  \bibinfo {address} {USA},\ \bibinfo {year} {2007})\BibitemShut {NoStop}%
\bibitem [{\citenamefont {Alonso}\ and\ \citenamefont
  {de~Vega}(2005)}]{alonso2005-multiple}%
  \BibitemOpen
  \bibfield  {author} {\bibinfo {author} {\bibfnamefont {D.}~\bibnamefont
  {Alonso}}\ and\ \bibinfo {author} {\bibfnamefont {I.}~\bibnamefont
  {de~Vega}},\ }\href {\doibase 10.1103/physrevlett.94.200403} {\bibfield
  {journal} {\bibinfo  {journal} {Physical Review Letters}\ }\textbf {\bibinfo
  {volume} {94}},\ \bibinfo {pages} {200403} (\bibinfo {year}
  {2005})}\BibitemShut {NoStop}%
\bibitem [{\citenamefont {Debnath}\ \emph {et~al.}(1989)\citenamefont
  {Debnath}, \citenamefont {Zhou},\ and\ \citenamefont
  {Moss}}]{debnath1989-remarks}%
  \BibitemOpen
  \bibfield  {author} {\bibinfo {author} {\bibfnamefont {G.}~\bibnamefont
  {Debnath}}, \bibinfo {author} {\bibfnamefont {T.}~\bibnamefont {Zhou}}, \
  and\ \bibinfo {author} {\bibfnamefont {F.}~\bibnamefont {Moss}},\ }\href
  {\doibase 10.1103/physreva.39.4323} {\bibfield  {journal} {\bibinfo
  {journal} {Physical Review A}\ }\textbf {\bibinfo {volume} {39}},\ \bibinfo
  {pages} {4323} (\bibinfo {year} {1989})}\BibitemShut {NoStop}%
\end{thebibliography}
\end{document}